\documentclass[letterpaper,english,reprint,nofootinbib,aps,superscriptaddress,showpacs,showkeys,prappl]{revtex4-1}

\usepackage{babel,calc,amsmath,amsthm,amssymb,graphicx,subfigure,xcolor,comment}
\usepackage{txfonts}
\usepackage{mathdots}
\usepackage{todonotes}
\usepackage[T1]{fontenc}
\usepackage[unicode=true]{hyperref}
\usepackage{booktabs}
\usepackage{threeparttable}
\usepackage{times}
\usepackage{braket}
\usepackage[utf8]{inputenc}
\hypersetup{
	colorlinks=true,       		
	linkcolor=blue,          	
	citecolor=blue,            
	urlcolor=black,           	
}

\begin{document}

\title{On-Demand Tailoring of High-Dimensional Orbital Angular Momentum States via a Phase-Only Hologram}

\author{Hao Sheng}
 \affiliation{
National Laboratory of Solid State Microstructures and School of Physics, Nanjing University, Nanjing 210093, China}
 \affiliation{Collaborative Innovation Center of Advanced Microstructures, Nanjing 210093, China}

\author{Zi-Shuo Wang}
 \affiliation{
National Laboratory of Solid State Microstructures and School of Physics, Nanjing University, Nanjing 210093, China}
 \affiliation{Collaborative Innovation Center of Advanced Microstructures, Nanjing 210093, China}

\author{Sheng-Qiang Jia}
 \affiliation{
National Laboratory of Solid State Microstructures and School of Physics, Nanjing University, Nanjing 210093, China}
 \affiliation{Collaborative Innovation Center of Advanced Microstructures, Nanjing 210093, China}

\author{Zi-Mo Cheng}
 \affiliation{
National Laboratory of Solid State Microstructures and School of Physics, Nanjing University, Nanjing 210093, China}
 \affiliation{Collaborative Innovation Center of Advanced Microstructures, Nanjing 210093, China}

\author{Li Fan}
 \affiliation{
College of Physics Science and Technology, Institute of Applied Photonic Technology, Yangzhou University, Yangzhou 225002, China}

\author{Jianping Ding}
 \affiliation{
National Laboratory of Solid State Microstructures and School of Physics, Nanjing University, Nanjing 210093, China}
 \affiliation{Collaborative Innovation Center of Advanced Microstructures, Nanjing 210093, China}

\author{Jia-Min Xu}
\email{jiaminxu@nju.edu.cn}
 \affiliation{
National Laboratory of Solid State Microstructures and School of Physics, Nanjing University, Nanjing 210093, China}
 \affiliation{Collaborative Innovation Center of Advanced Microstructures, Nanjing 210093, China}

\author{Zhi-Cheng Ren}
\email{zcren@nju.edu.cn}
 \affiliation{
National Laboratory of Solid State Microstructures and School of Physics, Nanjing University, Nanjing 210093, China}
 \affiliation{Collaborative Innovation Center of Advanced Microstructures, Nanjing 210093, China}

\author{Xi-Lin Wang}
\email{xilinwang@nju.edu.cn}
 \affiliation{
National Laboratory of Solid State Microstructures and School of Physics, Nanjing University, Nanjing 210093, China}
 \affiliation{Collaborative Innovation Center of Advanced Microstructures, Nanjing 210093, China}
 \affiliation{Jiangsu Physical Science Research Center, Nanjing 210093, China}
 \affiliation{Jiangsu Key Laboratory of Quantum Information Science and Technology, Nanjing University, Nanjing 210093, China}
 \affiliation{Hefei National Laboratory, Hefei 230088, China}
 \affiliation{Synergetic Innovation Center of Quantum Information and Quantum Physics, University of Science and Technology of China, Hefei 230026, China}

\author{Hui-Tian Wang}
 \affiliation{
National Laboratory of Solid State Microstructures and School of Physics, Nanjing University, Nanjing 210093, China}
 \affiliation{Collaborative Innovation Center of Advanced Microstructures, Nanjing 210093, China}
 \affiliation{Collaborative Innovation Center of Extreme Optics, Shanxi University, Taiyuan 030006, China}

\date{\today}

\begin{abstract}
Photonic orbital angular momentum (OAM) provides a naturally discrete and unbounded Hilbert space for high-dimensional encoding and has emerged as a powerful degree of freedom for advanced information processing. On-demand tailoring of both the amplitude and phase profiles of high-dimensional OAM states is essential for numerous applications, particularly in quantum information science. Here, we propose and experimentally demonstrate an efficient method for achieving such on-demand tailoring using a purely phase-only modulation scheme. The measured fidelity remains above 80\% even for a 128-dimensional state, demonstrating the scalability of the approach. Compared with previous techniques, our method achieves significantly higher generation efficiency, and this advantage becomes more pronounced as the dimensionality increases. Moreover, our approach substantially reduces experimental complexity while maintaining precise control, thereby offering a practical and scalable solution for high-dimensional OAM state engineering in both classical and quantum information processing.
\end{abstract}

\maketitle

\section{\label{sec:introduction}Introduction}
The orbital angular momentum (OAM) of light, originating from its spatially helical phase distribution~\cite{Allen1992}, has become a powerful degree of freedom in modern photonics. It provides a theoretically unbounded set of orthogonal computational basis states, enabling the construction of an infinite-dimensional Hilbert space. This unique property enables OAM-based multiplexing in classical optical communications~\cite{Wang2012, Bozinovic2013} and supports high-dimensional (HD) quantum information processing~\cite{Mair2001, Pires2010, Bavaresco2018, Erhard2018, Kong2020, Li2023, Karan2025, Sheng2025, Krenn2014, Qu2022}, offering notable advantages such as enhanced channel capacity~\cite{Martin2017, Hu2018, Hu2020}, improved resilience to noise~\cite{Cerf2002, Doda2021, Srivastav2022}, and simplified implementation of quantum logic operations~\cite{Barenco1995, Lanyon2009, Kiktenk2025}. Nevertheless, a key prerequisite for translating these theoretical advantages into practice is the capability to manipulate high-dimensional OAM states with both high precision and high efficiency, typically achieved via phase-only modulation.

A high-dimensional OAM state can be expressed as $|\Psi\rangle = \sum_m A_m e^{j\delta_m} |m\rangle$, where $|m\rangle$ denotes the OAM eigenstate of order $m$, and $A_m$ and $\delta_m$ represent the corresponding amplitude and phase, respectively. In classical optical communications and OAM-based multichannel applications, precise control of the amplitudes $\{A_m\}$ within a high-dimensional state, commonly referred to as OAM spectrum manipulation, is typically required. Phase-only holograms provide a physically feasible means of tailoring amplitudes; various schemes have been developed, such as iterative algorithms assisted by pattern search~\cite{Zhu2015}, azimuthal binary phase modulation~\cite{Fu2022}, two-fold symmetric multi-spiral-phase holography~\cite{Ren2024}, and deep-learning-assisted approaches~\cite{Zhou2023}.

In quantum science, state coherence plays a vital role in the construction of quantum superposition states, which is essential for unlocking quantum advantages. Consequently, practical quantum applications require not only the tailoring of the amplitude $A_m$ but also the control of the phase $\delta_m$ within the high-dimensional state. However, simultaneous customization of both amplitude and phase using phase-only holography remains challenging because of the limited capability of direct pure phase modulation. To enhance the amplitude tailoring capability, one approach employs blazed grating modulation, in which adjusting the depth of the blazed grating enables the hologram to selectively direct the desired portion of the beam into the first-order diffraction while routing the remainder into the zeroth or higher orders~\cite{Leach2005, Bolduc2013}. The diffraction efficiency depends on the phase depth---the closer the phase variation is to a full $2\pi$, the higher the light yield. An alternative approach is multiple-pixel encoding, which uses two spatially sampled phase elements~\cite{Mendoza2014} or a macro-pixel~\cite{Chen2015} to achieve simultaneous control of phase and amplitude. Nevertheless, the additional modulation imposed in the hologram inevitably limits the overall efficiency. Therefore, developing an efficient method for the on-demand tailoring of both amplitude and phase is of great importance for applications involving high-dimensional OAM states.

Here we propose and experimentally demonstrate a novel method for on-demand tailoring of both amplitude $A_m$ and phase $\delta_m$ in high-dimensional OAM states without introducing additional modulation, which in turn improves efficiency compared to methods that rely on extra modulation degrees. We design our phase-only hologram in an expanded space with more dimensions, ensuring that only the state in the desired dimensions exactly matches our target. Leveraging the orthogonality between computational states across all dimensions, the components in undesired dimensions do not affect the state in the intended subspace. Our hologram enables the conversion of a high-dimensional OAM state from a fundamental Gaussian beam using only a single spatial light modulator (SLM), thereby eliminating the need for complex cascaded modulation stages or multi-pass configurations. This significantly reduces experimental complexity while preserving precise control over the state structure. Compared with previous methods, our approach achieves a twofold enhancement in conversion efficiency for ten-dimensional and higher-dimensional states.

To validate the quality of our generated states, we perform projective measurements and reconstruct the density matrices via quantum state tomography. The experimental results demonstrate exceptional performance across a wide range of dimensions. For low-dimensional states, we achieve fidelities of 98.4\% for 4-dimensional states and 97.9\% for 8-dimensional states. Remarkably, even as the dimension scales significantly, the fidelity remains high, reaching 96.5\% for 16-dimensional states, 95.2\% for 32-dimensional states, and 92.0\% for 50-dimensional states. We also present an equal-amplitude 128-dimensional OAM state comprising the modes $m=-64,\ldots,-1,1,\ldots,64$, with randomly chosen phases and a fidelity of 82.0\% [Fig.~\ref{fig:3}(e)]. These results illustrate the feasibility of extending the method to higher-dimensional mode sets.

\section{\label{sec:principles}Principles}

\subsection{On-Demand Tailoring of Amplitude and Phase}

\begin{figure}[ht]
	\centering
	\includegraphics[width=.9\linewidth]{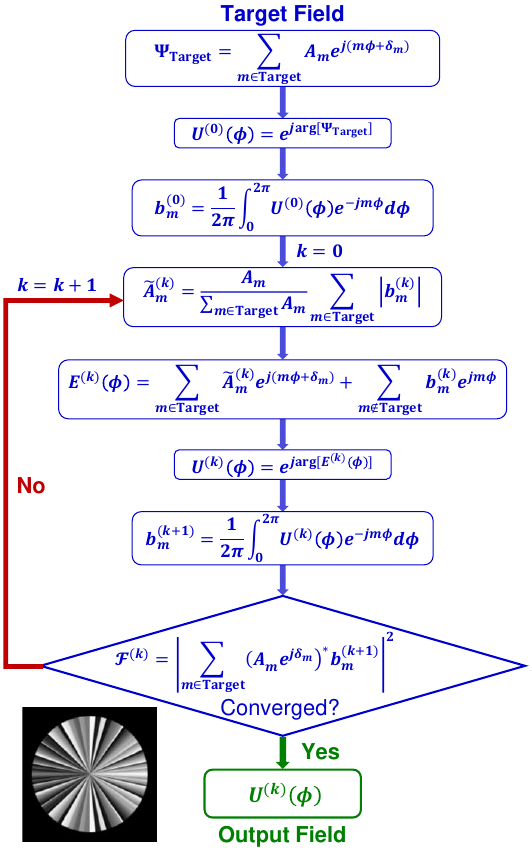}
	\caption{Fast-converging pure spiral phase algorithm for simultaneous amplitude and phase modulation.}
	\label{fig:1}
\end{figure}

For a high-dimensional OAM state of $\sum_m A_m e^{j\delta_m} | m \rangle$, the corresponding optical field---referred to as the target field---can be expressed in complex form as
\begin{equation}
\Psi_{\text{Target}} = \sum_{m\in\text{Target}} A_m e^{j(m\phi + \delta_m)},
\end{equation}
where $\phi$ is the azimuthal angle in the polar coordinate system, and $A_m \ge 0$ and $\delta_m$ denote the target amplitude and phase of the OAM mode $m$, respectively. The coefficients satisfy the normalization condition $\sum_m A_m^2 = 1$. Our goal is to prepare the target field $\Psi_{\text{Target}}$ using a phase-only hologram. To this end, we propose an iterative algorithm that simultaneously optimizes the amplitude and phase, as shown in Fig.~\ref{fig:1}, enabling rapid convergence toward the optical field $\Psi_{\text{Target}}$ that corresponds to the desired OAM superposition state.

Our approach begins to extract only the phase information of the target field, temporarily neglecting its amplitude. The initial phase distribution is defined as
\begin{equation}
\Phi^{(0)}(\phi) = \arg \bigl[ \Psi_{\text{Target}} \bigr],
\end{equation}
and the corresponding phase-only field is
\begin{equation}
U^{(0)}(\phi) = e^{j\Phi^{(0)}(\phi)}.
\end{equation}

At each iteration $k$, we decompose $U^{(k)}(\phi)$ into its OAM modes via
\begin{equation}
b_m^{(k)} = \frac{1}{2\pi} \int_0^{2\pi} U^{(k)}(\phi) e^{-jm\phi} \, d\phi.
\label{eq:bm}
\end{equation}

To construct a modified phase distribution, we form a new complex field combining target and non-target modes. For non-target modes, we retain $b_m^{(k)}$ and weight it by $e^{jm\phi}$. For target modes, we replace the coefficient with an updated amplitude
\begin{equation}
\tilde{A}_m^{(k)} = \frac{A_m}{\sum_{m \in \text{Target}} A_m} \sum_{m \in \text{Target}} \bigl| b_m^{(k)} \bigr|,
\end{equation}
where the first term normalizes the target amplitudes and the second sums the moduli of the current target-mode coefficients. The updated field is
\begin{equation}
\begin{split}
E^{(k)}(\phi) &= \sum_{m \in \text{Target}} \tilde{A}_m^{(k)} e^{j(m\phi + \delta_m)} \\
&\quad + \sum_{m \notin \text{Target}} b_m^{(k)} e^{jm\phi}.
\end{split}
\end{equation}

The next-iteration phase is extracted as
\begin{equation}
\Phi^{(k)}(\phi) = \arg \bigl[ E^{(k)}(\phi) \bigr].
\end{equation}

Evaluating the fidelity between the generated field and the target state as
\begin{equation}
\mathcal{F}^{(k)} = \left| \sum_{m \in \text{Target}} \bigl( A_m e^{j\delta_m} \bigr)^* b_m^{(k+1)} \right|^2.
\label{eq:fidelity}
\end{equation}

The iteration cycles through Eqs.~\eqref{eq:bm}-\eqref{eq:fidelity} until $\mathcal{F}^{(k)}$ converges to a stable maximum, typically within a few iterations. The final phase $\Phi^{(\text{final})}(\phi)$ is then encoded onto a phase-only SLM for high-fidelity experimental generation of the target high-dimensional OAM state.

\begin{figure}[!htb]
	\centering
	\includegraphics[width=.9\linewidth]{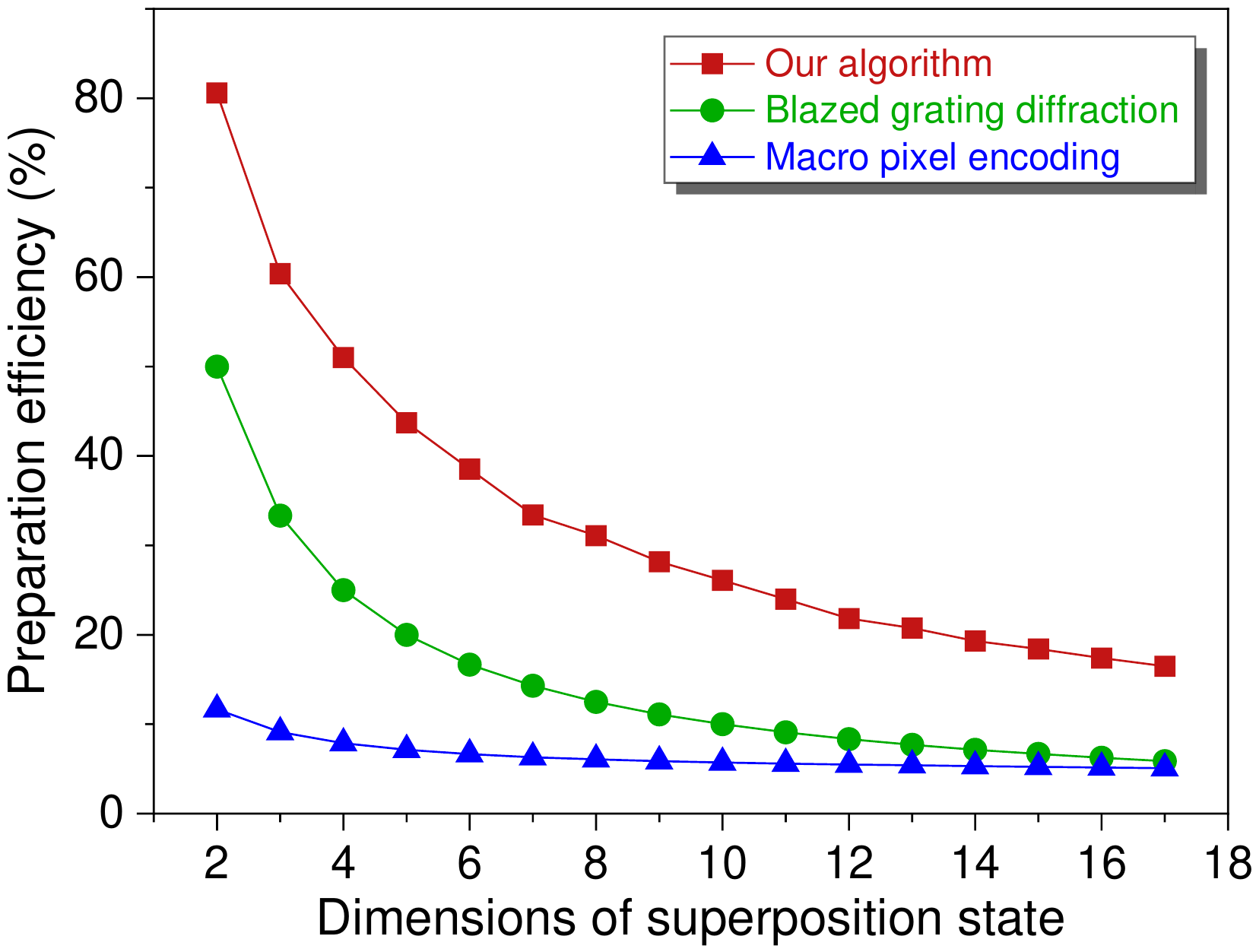}
	\caption{Comparison of the generation efficiencies among three different algorithms for an equally superposed high-dimensional OAM state with zero relative phase among the constituent modes.}
	\label{fig:2}
\end{figure}

To rigorously assess the performance of our algorithm for generating high-dimensional OAM states, we compare its efficiency with two existing methods: blazed grating modulation~\cite{Leach2005, Bolduc2013} and multiple-pixel encoding~\cite{Mendoza2014, Chen2015}. All three methods can generate arbitrary high-dimensional OAM states by encoding complex amplitude and phase onto a single phase-only hologram. As shown in Fig.~\ref{fig:2}, for an $N$-dimensional OAM state with uniform amplitudes and zero relative phases ($\delta_m=0$), our algorithm consistently achieves higher generation efficiency, particularly for $N \ge 10$, where it exhibits a twofold enhancement. These results clearly demonstrate the advantage of our specialized algorithm over previous methods for high-dimensional OAM state generation.

\subsection{Quantum State Tomography and Fidelity Evaluation}

To evaluate the quality of the generated high-dimensional OAM states, we perform quantum state tomography (QST) based on a set of projection measurements~\cite{Wootters1989, James2001}. The experimental data are used to reconstruct the density matrix $\rho$ via maximum-likelihood estimation (MLE)~\cite{Hradil2004}. The fidelity between the reconstructed state and the ideal target state $|\Psi_{\text{Target}}\rangle$ is then computed as a measure of the generation performance.

A second spatial light modulator (SLM2) is used to sequentially project the generated state onto a complete set of measurement bases. Within a $d$-dimensional OAM subspace spanned by $\{|m\rangle\}$ $(m=m_0, \cdots, m_0+d-1)$, we conduct $d^2$ projection measurements that form a tomographically complete set. These consist of:
\begin{itemize}
	\item Projections onto each of the $d$ OAM eigenstates $|m\rangle$ are implemented using forked holograms that convert a specific OAM mode $|m\rangle$ into a Gaussian beam, which is subsequently coupled into a single-mode fiber.
	\item Projections onto superposition states $\frac{1}{\sqrt{2}}(|m\rangle + |m'\rangle)$ and $\frac{1}{\sqrt{2}}(|m\rangle + j|m'\rangle)$ for every pair of distinct modes $(|m\rangle,|m'\rangle)$ with $m < m'$. These $d(d-1)$ measurements are realized through phase holograms that combine two OAM components with controlled relative phases \cite{Dada2011}.
\end{itemize}

Let $\{|\psi_i\rangle\}$ ($i=1, \cdots, d^2$) represent the set of projection states, involving $|m\rangle$, $\frac{1}{\sqrt{2}}(|m\rangle + |m'\rangle)$, and $\frac{1}{\sqrt{2}}(|m\rangle + j|m'\rangle)$ with $m < m'$, and the corresponding projectors are $|\psi_i\rangle\langle\psi_i|$. For each projection $i$, the optical power $P_i$ transmitted through the setup is recorded by a power meter. Assuming constant input power and detection efficiency across all measurements, the normalized power $p_i = P_i / \sum_{i} P_{i}$ gives the experimental probability, which ideally satisfies $p_i = \operatorname{Tr}(\rho |\psi_i\rangle\langle\psi_i|)$.

Experimental noise and statistical errors often cause the direct inversion of the raw probabilities $p_i$ to yield an unphysical density matrix, for example, one that is not positive semidefinite or does not have unit trace. To address this issue appropriately, we reconstruct $\rho$ via maximum-likelihood estimation, which explicitly enforces the required physical constraints~\cite{James2001,Hradil2004}. For a candidate physical state $\hat{\rho}$, the likelihood given the data is
\begin{equation}
\mathcal{L}(\hat{\rho}) \propto \prod_{i=1}^{d^2} \exp\left[ -\frac{\bigl( N\langle\psi_i|\hat{\rho}|\psi_i\rangle - n_i \bigr)^2}{2N\langle\psi_i|\hat{\rho}|\psi_i\rangle} \right],
\end{equation}
with $N$ being the total detected photons and $n_i = Np_i$ the count for the $i$-th projection. The log-likelihood is then
\begin{equation}
\ln \mathcal{L}(\hat{\rho}) = \text{const} - \sum_{i=1}^{d^2} \frac{\bigl( N\langle\psi_i|\hat{\rho}|\psi_i\rangle - n_i \bigr)^2}{2N\langle\psi_i|\hat{\rho}|\psi_i\rangle}.
\end{equation}

The MLE procedure seeks $\hat{\rho}_{\text{MLE}}$ that maximizes $\mathcal{L}(\hat{\rho})$ (or minimizes the negative log-likelihood) subject to $\hat{\rho} \geq 0$ and $\operatorname{Tr}(\hat{\rho}) = 1$. This optimization is typically solved via iterative methods, including expectation-maximization and gradient-based convex algorithms \cite{James2001}.

With the reconstructed physical state $\hat{\rho}_{\text{MLE}}$ obtained from the iterative optimization, the fidelity to the target state $|\Psi_{\text{Target}}\rangle$ is given by
\begin{equation}
\mathcal{F} = \langle \Psi_{\text{Target}} | \hat{\rho}_{\text{MLE}} | \Psi_{\text{Target}} \rangle.
\end{equation}
This fidelity directly reflects the overlap between the prepared state and the target superposition state.

\section{\label{sec:experiment}Experimental Verification}

In the experiment, a fundamental Gaussian laser at a wavelength of $\lambda = 1064~\text{nm}$ is expanded and collimated using a $4f$ telescope to uniformly illuminate the active area of SLM1. This SLM1 displays the on-axis phase hologram generated by our iterative algorithm. Upon reflection, the beam acquires the prescribed helical phase profile, thereby generating the desired arbitrary superposition OAM state. A subsequent $4f$ imaging system transfers the field from SLM1 to SLM2, which completes the projection measurement together with a single-mode fiber coupled to the power meter. The measured power values, together with the known projection operators, are then used to reconstruct the density matrix of the generated state via maximum-likelihood estimation. This reconstruction yields a physical density matrix that provides the best fit to the experimental data. Finally, the fidelity between the reconstructed state and the ideal target state is calculated.

\begin{figure*}[!htbp]
	\centering
	\includegraphics[width=0.9\linewidth]{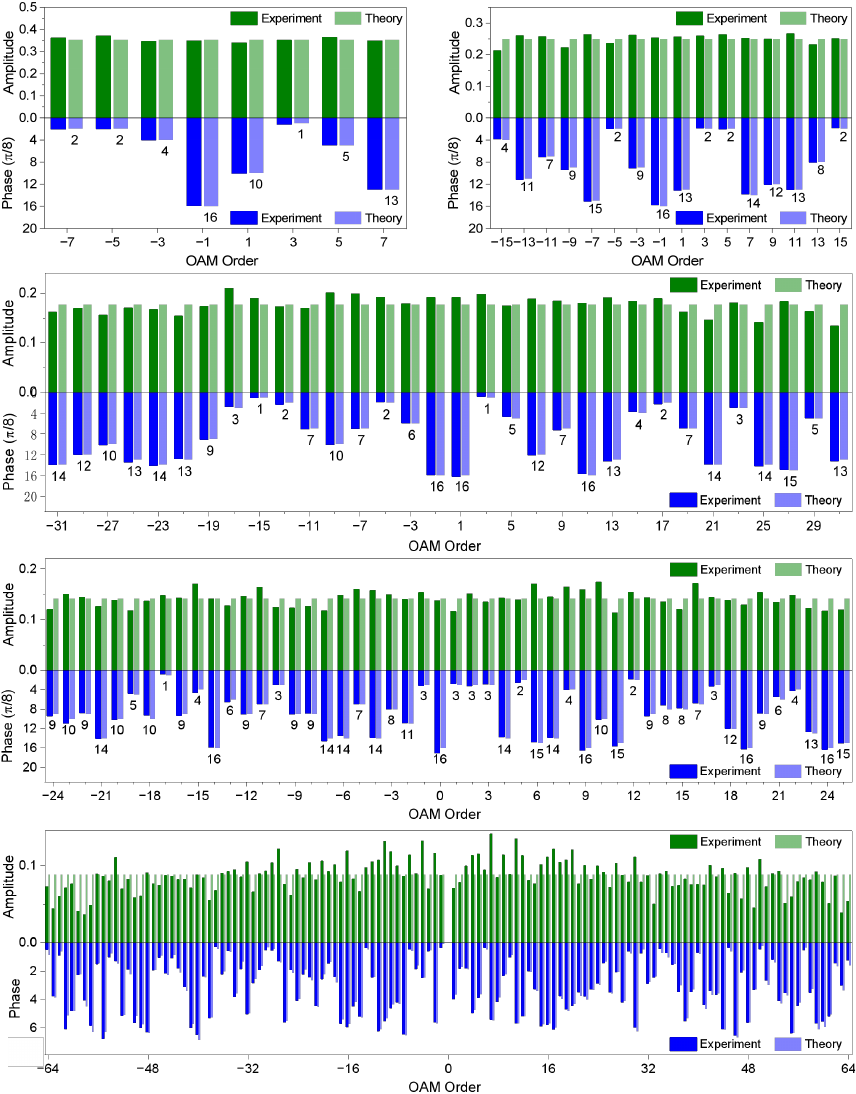}%
	\caption{Results for equal-amplitude high-dimensional OAM superposition states. (a) 8D odd-index; (b) 16D odd-index; (c) 32D odd-index; (d) 50D consecutive-index; (e) 128D, with $m=-64,\ldots,-1,1,\ldots,64$. For (a)--(d), phases $\delta_m$ are randomly chosen from the discrete set $n\pi/8$; the numbers in the lower panels indicate the corresponding values of $n$. In (e), the phases are randomly chosen and are not restricted to integer multiples of $\pi/8$; the fidelity is 82.0\%.}
	\label{fig:3}
\end{figure*}

\begin{figure*}[!htbp]
	\centering
	\includegraphics[width=.9\linewidth]{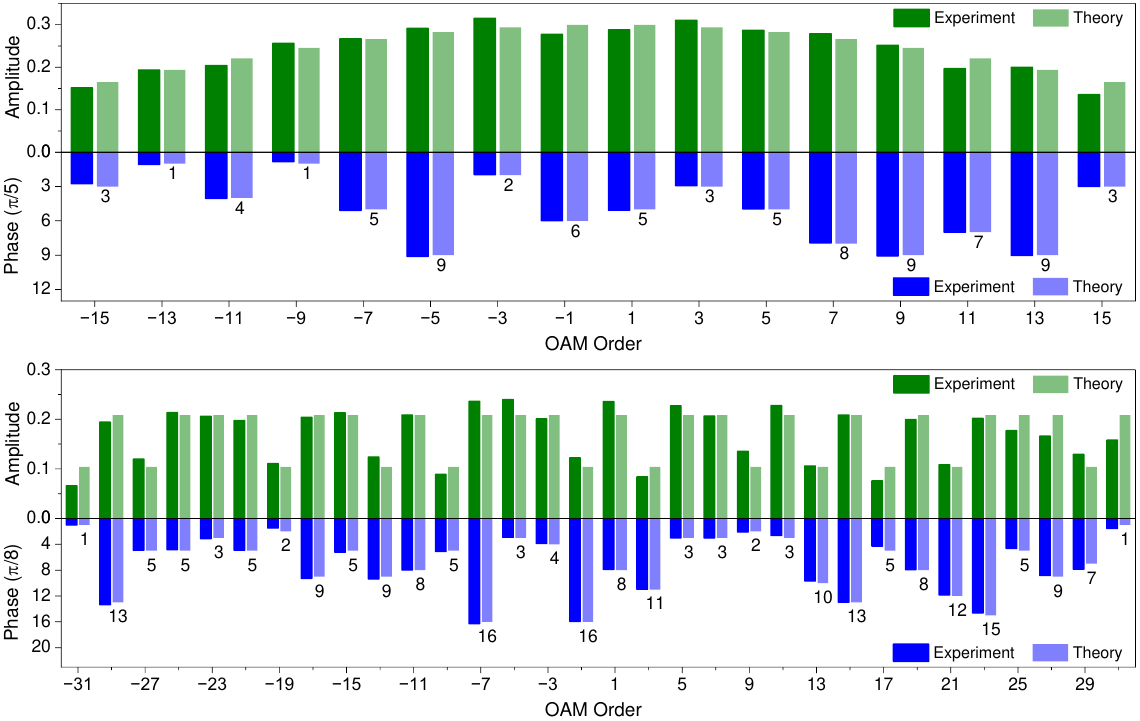}%
	\caption{Results for high-dimensional OAM states with arbitrary amplitude and phase encoding. (a) sin/cos amplitude distribution and phase distribution of $\delta_m = n\pi/5$, where $n$ takes in turn the first 16 digits of irrational number $\pi$ ($3,1,4,1,5,9,2,6,5,3,5,8,9,7,9,3$); (b) random amplitudes (1 or 2) and random phases, where phases $\delta_m$ are randomly chosen from the discrete set $n\pi/8$. The numbers in the lower panels of each figure indicate the corresponding values of $n$.}
	\label{fig:4}
\end{figure*}

First, we demonstrate a simple case of OAM comb tailoring, specifically a four-dimensional OAM superposition state with equal amplitudes ($A_m \equiv 0.5$) and equal phases ($\delta_m \equiv 0$) across all constituent modes. Quantum state tomography is performed using a complete set of $d^2 = 16$ projection measurements, and the density matrix is reconstructed via maximum-likelihood estimation. From the reconstructed density matrix, we directly extract the amplitude and phase distributions of the generated OAM superposition state. The experimental values were $(A_1, A_2, A_3, A_4) = (0.557, 0.477, 0.493, 0.469)$ and $(\delta_1, \delta_2, \delta_3, \delta_4) = (0.043 \pi, 0.065 \pi, -0.094 \pi, -0.014 \pi)$, corresponding to a fidelity of 98.4\% between the reconstructed state and the ideal target state, which confirms the high quality of the generated state.

The theoretical generation efficiency, defined as the ratio of the power residing in the desired OAM modes to the total power, is calculated to be 51.0\%. Experimentally, we determine the generation efficiency as the ratio of the total power carried by all target OAM modes to the total power incident on the fiber coupler. To quantify this, we first measure the total power before the fiber coupler using a power meter. Subsequently, we sequentially convert each target OAM mode into the fundamental Gaussian mode by displaying the corresponding phase-flattening holograms on SLM2 and measure the power coupled into the single-mode fiber. The sum of these measured powers over all target modes yields the total power contained in the desired modal components. The experimentally obtained efficiency is 39.0\%. Considering an overall coupling efficiency of approximately 80.0\% (including fiber-coupling losses, imperfect mode matching, and Fresnel reflections), the experimental efficiency is in good agreement with the theoretical prediction.

Second, we extend our method to higher dimensions and more complex phase distributions. A primary motivation is to evaluate the scalability of our algorithm as the Hilbert space dimension increases, since practical applications often require operation in high-dimensional regimes. We first consider a series of OAM states with equal-amplitude coefficients and randomly chosen phases. The equal-amplitude condition provides a clear benchmark for evaluating the algorithm's ability to maintain uniform modal weights across all targeted OAM modes, while the random phases test its robustness against arbitrary phase configurations lacking any special symmetry or pattern. For the 8-, 16-, 32-, and 50-dimensional states, the phases $\delta_m$ are randomly chosen from the discrete set $n\pi/8$ (with $n = 1, 2, \cdots, 16$), corresponding to 16 equally spaced phase levels over the range $[0, 2\pi)$. For the 128-dimensional state, the phases are randomly chosen without being restricted to integer multiples of $\pi/8$. The following cases are considered:

\begin{itemize}
	\item An 8-dimensional state restricted to odd OAM indices, $m = -7,-5,-3,-1,1,3,5,7$, yielding a reconstructed fidelity of 97.9\% [Fig.~\ref{fig:3}(a)].
	\item A 16-dimensional state with odd indices, $m = -15,-13,\cdots,13,15$, achieving a fidelity of 96.5\% [Fig.~\ref{fig:3}(b)].
	\item A 32-dimensional state with odd indices, $m = -31,-29,\cdots,29,31$, with a fidelity of 95.2\% [Fig.~\ref{fig:3}(c)].
	\item A 50-dimensional state spanning $m = -24,-23,\cdots,25$, for which the fidelity is measured to be 92.0\% [Fig.~\ref{fig:3}(d)].
	\item A 128-dimensional state spanning $m = -64,\ldots,-1,1,\ldots,64$, with a fidelity of 82.0\% [Fig.~\ref{fig:3}(e)].
\end{itemize}

These results illustrate the range of mode sets considered, including a 128-dimensional equal-amplitude spectrum.

Next, we explore high-dimensional OAM superposition states with non-uniform amplitude profiles and arbitrarily selected phases, demonstrating the flexibility of our method. Two representative examples are shown in Fig.~\ref{fig:4}.

\begin{itemize}
	\item \textbf{The sin/cos amplitude distribution with $\pi$-digit phase encoding}: For a 16-dimensional state confined to the odd-index subspace ($m = -15,-13,\cdots,13,15$), the target amplitudes follow a sin/cos distribution given by $A_m \propto 1 + \cos(m \pi /32)$, appropriately normalized. The phases are encoded using the first 16 digits of irrational number $\pi$ (i.e., $3,1,4,1,5,9,2,6,5,3,5,8,9,7,9,3$) in units of $\pi/5$. The fidelity of the generated state reaches 95.9\% [Fig.~\ref{fig:4}(a)].

	\item \textbf{The random amplitudes and phases}: For a 32-dimensional odd-index state ($m = -31,-29,\cdots,29,31$), the amplitudes are randomly assigned values of either 1 or 2 (before normalization), and the phases are randomly selected from the discrete set $n\pi/8$ ($n$ takes any integer between 1 and 16). The resulting fidelity is 94.0\% [Fig.~\ref{fig:4}(b)].
\end{itemize}

These experiments confirm that our iterative algorithm can successfully encode arbitrary combinations of amplitudes and phases into a single phase-only hologram, thereby enabling the generation of virtually any desired high-dimensional OAM state.

\section{\label{sec:conclusion}Conclusion}

We have developed and experimentally validated an iterative algorithm that encodes both the amplitude and phase information of an arbitrary high-dimensional OAM superposition state into a single phase-only hologram. This single-hologram approach eliminates the need for cascaded modulation stages, thereby simplifying the experimental setup while maintaining high precision. Using one SLM for state preparation and a second SLM for projection measurements, we perform quantum state tomography and reconstruct the density matrix via maximum-likelihood estimation. The experimental results demonstrate high performance across a broad range of dimensions: a fidelity of 98.4\% is obtained for a four-dimensional state and 92.0\% for a 50-dimensional state. For an equal-amplitude 128-dimensional OAM state with randomly chosen phases, the fidelity is 82.0\%. Moreover, the generation efficiency of our algorithm surpasses that of conventional complex-amplitude methods, and this advantage becomes increasingly pronounced as the dimensionality grows.

Our efficient approach for on-demand tailoring of both amplitude and phase in high-dimensional OAM states may find important applications in both state preparation~\cite{Dong2023} and state measurement~\cite{Dada2011}. Particularly in the measurement context, our approach---combining a phase-only hologram with a single-mode fiber---provides an efficient means to implement arbitrary high-dimensional OAM state projection measurements. The efficiency enhancement scales exponentially with the number of photons, making our approach increasingly valuable for quantum applications in multi-photon high-dimensional quantum information processing~\cite{Malik2016, Erhard2018-2, Liu2026}.

\begin{acknowledgments}
This work was supported by Quantum Science and Technology-National Science and Technology Major Project (Grant No. 2021ZD0301400); National Natural Science Foundation of China (Grants No. 12427808, 12234009, 12274215, 12404382, U25A6008 and U25A20194); Fundamental Research Funds for the Central Universities (Grant No. 2026300374); Program for Innovative Talents and Entrepreneurs in Jiangsu; Key R\&D Program of Jiangsu Province (Grant No. BE2023002); Natural Science Foundation of Jiangsu Province (Grant No. BK20233001).

\end{acknowledgments}

\begin{center}
\textbf{AUTHOR DECLARATIONS}
\end{center}

\noindent
\textbf{Conflict of Interest}

The authors have no conflicts to disclose.

\noindent
\textbf{Author Contributions}

\noindent
\textbf{Hao Sheng}: Conceptualization (lead); Data curation (lead); Formal analysis (lead); Investigation (lead); Methodology (lead); Visualization (equal); Writing -- original draft (lead).
\textbf{Zi-Shuo Wang}: Data curation (equal); Formal analysis (equal); Writing -- review \& editing (equal).
\textbf{Sheng-Qiang Jia}: Data curation (equal); Formal analysis (equal); Writing -- review \& editing (equal).
\textbf{Zi-Mo Cheng}: Data curation (equal); Formal analysis (equal).
\textbf{Li Fan}: Data curation (supporting).
\textbf{Jianping Ding}: Data curation (equal); Writing -- original draft (equal); Writing -- review \& editing (equal).
\textbf{Jia-Min Xu}: Conceptualization (equal); Data curation (equal); Formal analysis (equal); Funding acquisition (equal); Methodology (equal); Project administration (lead); Writing -- original draft (equal).
\textbf{Zhi-Cheng Ren}: Conceptualization (equal); Data curation (equal); Formal analysis (equal); Funding acquisition (equal); Methodology (equal); Visualization (equal); Project administration (lead); Writing -- original draft (equal).
\textbf{Xi-Lin Wang}: Conceptualization (equal); Data curation (equal); Funding acquisition (lead); Project administration (lead); Writing -- original draft (equal); Writing -- review \& editing (lead).
\textbf{Hui-Tian Wang}: Conceptualization (equal); Data curation (equal); Formal analysis (equal); Funding acquisition (equal); Methodology (equal); Project administration (equal); Writing -- review \& editing (lead).

\begin{center}
\textbf{DATA AVAILABILITY}
\end{center}

The data that support the findings of this study are available from the corresponding author upon reasonable request.

\end{document}